\documentclass[12pt]{article}
\usepackage{graphicx}
\def\holt{Physics Division, Argonne National Laboratory, Argonne, Illinois 60439}
\def\brad{Kellogg Radiation Laboratory, California Institute of Technology, Pasadena, California 91125}
\def\support1{\footnote{This work is supported by the U.S. National Science Foundation under grant 1506459.}
\def\support2(\footnote{This work is supported by the U.S. Department of Energy (DOE), Office of Science, Office of Nuclear Physics, under contract No. DE-AC02-06CH11357.}}

\def\Title#1{\begin{center} {\Large #1 } \end{center}}
\def\Author#1{\begin{center}{ \sc #1} \end{center}}
\def\Address#1{\begin{center}{ \it #1} \end{center}}

\newenvironment{Presented}{\begin{quotation} \begin{center} 
             PRESENTED AT\end{center}\bigskip 
      \begin{center}\begin{large}}{\end{large}\end{center} \end{quotation}}
\def\Acknowledgements{\bigskip  \bigskip \begin{center} \begin{large}
             \bf ACKNOWLEDGEMENTS \end{large}\end{center}}
\def\beq{\begin{equation}}
\def\eeq#1{\label{#1}\end{equation}}
\def\eeqn{\end{equation}}

\def\beqa{\begin{eqnarray}}
\def\eeqa#1{\label{#1}\end{eqnarray}}
\def\eeqan{\end{eqnarray}}

\let\bar=\overbar

\def\Dslash{\not{\hbox{\kern-4pt $D$}}}
\def\dslash{\not{\hbox{\kern-2pt $\del$}}}

\def\msb{{\bar{\ssstyle M \kern -1pt S}}}

\begin{document}
\begin{titlepage}
%\pubblock
\vfill

\Title{Sensitivity study for the $^{12}$C($\alpha,\gamma$)$^{16}$O\\ astrophysical reaction rate}
\vfill
\Author{ R. J. Holt$^{1,2}$, B.W. Filippone$^2$, and Steven C. Pieper$^1$}
\Address{$^1$\holt}
\Address{$^2$\brad}
\vfill
\begin{abstract}
The $^{12}$C($\alpha,\gamma$)$^{16}$O reaction has a key role in nuclear astrophysics.  A multilevel R-matrix analysis was used to make extrapolations of the astrophysical S factor for this reaction to the stellar energy of 300~keV. The statistical precision of the S-factor extrapolation was determined by performing multiple fits to existing randomized E1 and E2 ground state data, according to experimental errors. The impact of a future proposed experiment at Jefferson Laboratory (JLab) was assessed within this framework.  The proposed JLab experiment will make use of a high-intensity low-energy bremsstrahlung beam that impinges on an oxygen-rich single-fluid bubble chamber in order to measure the total cross section for the inverse $^{16}$O($\gamma,\alpha$)$^{12}$C reaction.  The importance of low energy data as well as high precision data was investigated.
\end{abstract}
\vfill
\begin{Presented}
Thirteenth Conference on the Intersections of Particle and Nuclear Physics (CIPANP2018) \\Palm Springs, California, USA, May 29-June 3, 2018\\

Report No. CIPANP2018-Holt
\end{Presented}
\vfill
\end{titlepage}
\def\thefootnote{\fnsymbol{footnote}}
\setcounter{footnote}{0}

\section{Introduction}
The $^{12}$C($\alpha,\gamma$)$^{16}$O reaction is believed to be one of the most important reactions in nuclear astrophysics.  The purpose of this study is not to provide an improved R-matrix analysis of the data for this reaction, but rather to have a reasonable R-matrix fit that can be used as a basis for comparison to fits with and without projected JLab data in order to assess the possible role of the JLab data in reducing the overall uncertainty in the cross section.  An excellent R-matrix analysis of this reaction and review of the subject is given in ref.{\cite{deBoer:2017ldl}.  In the present work, the R-matrix was used to calculate the total cross section, $\sigma(E)$, for alpha-capture to the ground state.  The cross section was then used to calculate the astrophysical $S$ factor given by

\begin{equation}
S(E) = \sigma(E)Ee^{2\pi \eta}
\end{equation}
where $E$ is the energy in the center of mass and $\eta$ is the Sommerfeld parameter, $\sqrt{\frac{\mu}{2E}}Z_1Z_2\frac{e^2}{\hbar^2}$ and $\mu$ is the reduced mass of the carbon ion and alpha particle.  Measurements of the $S$ factor as a function of energy are often reported in the literature.  For the $^{12}$C($\alpha,\gamma$)$^{16}$O reaction, the value of $S$ at $E=300\ keV$ is typically quoted as this is the most probable energy for stellar helium burning.  Of course, the cross section is so small at 300~keV that it cannot be directly measured.  Thus, extrapolations to 300~keV must be performed to study the impact of data on the extrapolation.

\section{R-matrix approach}

According to the R-matrix theory, the nuclear wave function, $\Psi_{(E(J)}$, can be expanded in terms of a complete set of states, $X_{\lambda(J)}$.

\begin{equation}
\Psi_{E(J)} = i\hbar^{1/2}e^{-i\phi_c}\sum_{\lambda\mu}A_{\lambda\mu}\Gamma_{\lambda\mu}^{1/2}X_{\lambda(J)}  
\end{equation}
where $\phi_c$ is a Coulomb phase shift and $A_{\lambda\mu}$ is the matrix that relates the internal wave function and the observed resonances.

\begin{equation}
\left(A^{-1}\right)_{\lambda\mu}=\left(E_\lambda - E\right)\delta_{\lambda\mu} - \xi_{\lambda\mu}  
\end{equation}
where $E_\lambda$ is a level energy, $\delta_{\lambda\mu}$ is the Kronecker delta and $\xi$ is given in terms of the Coulomb shift factor, $S_c$, the boundary condition constant, $b_c$, and the Coulomb penetration factor, $P_c$, 

\begin{equation}
\xi_{\lambda\mu} = \sum_c [(S_c - b_c) + iP_c]\gamma_{\lambda c}\gamma_{\mu c} 
\end{equation}
where c refers to the available channels, in this case essentially the $\alpha$ channel only, and $\gamma_{\lambda c}$  are the reduced width amplitudes.

In this simple R-matrix analysis, we assume that only two channels are open: photon and alpha. 
The collision matrix for radiative capture of multipolarity $\mathcal{L}$ to the ground state is given by

\begin{equation}
  U_{\gamma \alpha}^{lJ\mathcal{L}} = i e^{-i\phi_l} \sum_{\lambda\mu} A_{\lambda\mu}\Gamma_{\lambda\alpha l J}^{1/2}\Gamma_{\mu\gamma l J}^{1/2} 
\end{equation}
where $\Gamma_{\lambda\alpha lJ}$ and $\Gamma_{\mu\gamma lJ}$ are the ground state $\alpha$ and radiative widths, respectively.
The cross section was then calculated from the collision matrix.

Only ground state transitions and excitations were taken into account in this analysis.  Furthermore, only statistical errors were considered in this study.  A channel radius of 5.43 fm was chosen to be consistent with a previous analysis\cite{deBoer:2017ldl}.  Five E1 resonance levels and four E2 resonance levels were employed in the internal part of the the R-matrix analysis.  This analysis is similar to that of ref{\cite{Azuma:2010zz}, ref{\cite{Holt:1978zz}, and the details comport with  Lane and Thomas\cite{Lane:1948zh}.   In order to speed up the computations, the external part was turned off in the present analysis.  This external contribution is most sensitive to the E2 part of the cross section since the E1 external part is greatly reduced by isospin symmetry.  In fact, the external E1 part would vanish under perfect isospin conservation.  The fit was performed for data less than 3 MeV where the external contribution is small for the E2 data.  As a check, the external piece was turned on for several fits, but did not significantly change the results.

\section{Fits and projections for S$_{E1}$, S$_{E2}$ and $S$ }

A SIMPLEX fitter was used for the present work.  The best R-matrix fit of the existing E1 and E2 S-factor data, shown in Fig.~\ref{fig1}, was taken as the most probable description of the S-factor data. In order to explore the statistical variation in the S-factor extrapolations, pseudo-data were then created by random variation according to a Gaussian probability distribution about the best fit S-factor values at the measured energies.  In the random variations, the individual pseudo-data uncertainties were set equal to the measured uncertainties multiplied by the square root of the ratio of the original best fit values to the original measured values.  For the subtheshold states the radiative widths were fixed at the measured values and the reduced alpha widths were allowed to vary.  The reduced alpha and radiative width of the first excited E1 state were allowed to vary in the fit, while the radiative width of the fifth E1 state was allowed to vary.  The radiative width of  the fourth E2 R-matrix level was also allowed to vary.
 The first excited E2 state is very narrow and the parameters of this level were fixed at those of ref.{\cite{deBoer:2017ldl}.  All other parameters were fixed at those of ref.{\cite{deBoer:2017ldl}.   Also, following ref. {\cite{deBoer:2017ldl}, the fits were performed by maximizing $L$ rather than minimizing $\chi^2$, where $L$ is given by

\begin{equation}
L = \sum_i ln[(1-exp(-R_i/2))/R_i]
\end{equation}
and $R_i=(f(x_i)-d_i)^2/\sigma_i^2$ is the usual quantity used in $\chi^2$ minimizations.  Here $f(x_i)$ is the function to be fitted to data, $d_i$, with statistical error $\sigma_i$.  The $L$ maximization has the feature that it reduces the impact of large error bar data on the fit and generally gives larger S-factor uncertainties in projected values of $S(300~ keV)$ than that of a $\chi^2$ minimization.

The parameters of the bound levels are very important for the projection to 300~keV. The resonance energies were fixed, but the parameters, $E_\lambda$, depend on the reduced width of the levels. The reduced widths of the bound states were allowed to vary, so the $E_\lambda$ varies.  The R-matrix boundary condition constants were fixed to cancel out this effect for the second levels so that $E_\lambda=E_R$ for these levels. For the third and higher levels, the reduced widths were not varied because alpha elastic scattering determined these widths and allowing them to vary did not make a significant difference. 
 
 The data sets used in the present fit are given in refs. \cite{Dyer:1974pgc,Kremer:1988zz,Redder:1987xba,Ouellet:1992zz,Roters:1999zz,Gialanella:2001ayx,Kunz:2001zz,Assuncao:2006vy,Makii:2009zz,Plag:2012zz} and the E1 and E2 ground state data are shown in Fig.\ref{fig1}.
  The parameters that were used in the R-matrix curve, shown in Fig.~\ref{fig1}, are given in table~\ref{two}.

\begin{figure}[ht]
\includegraphics[width=3.0in]{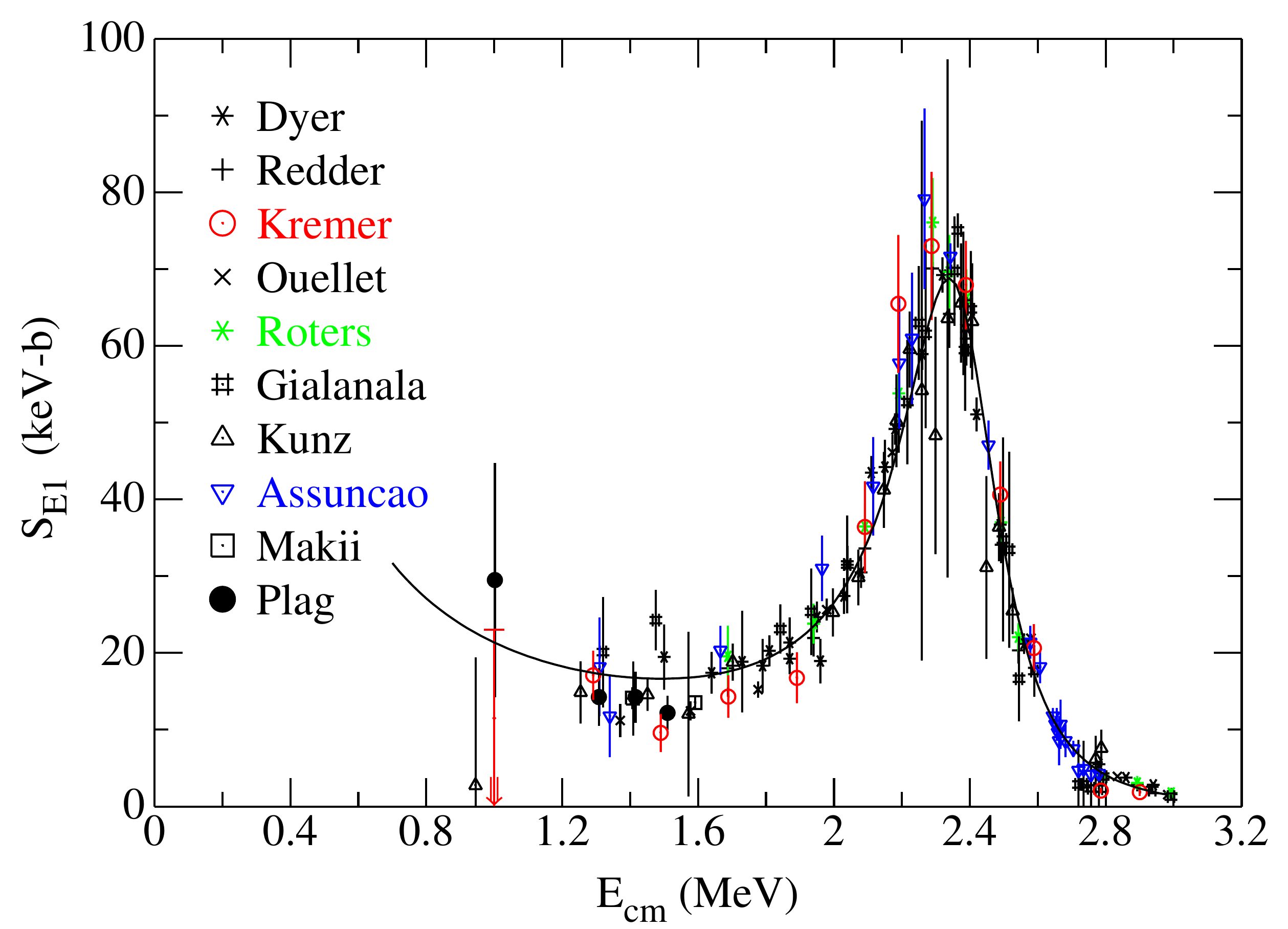}
\includegraphics[width=3.0in]{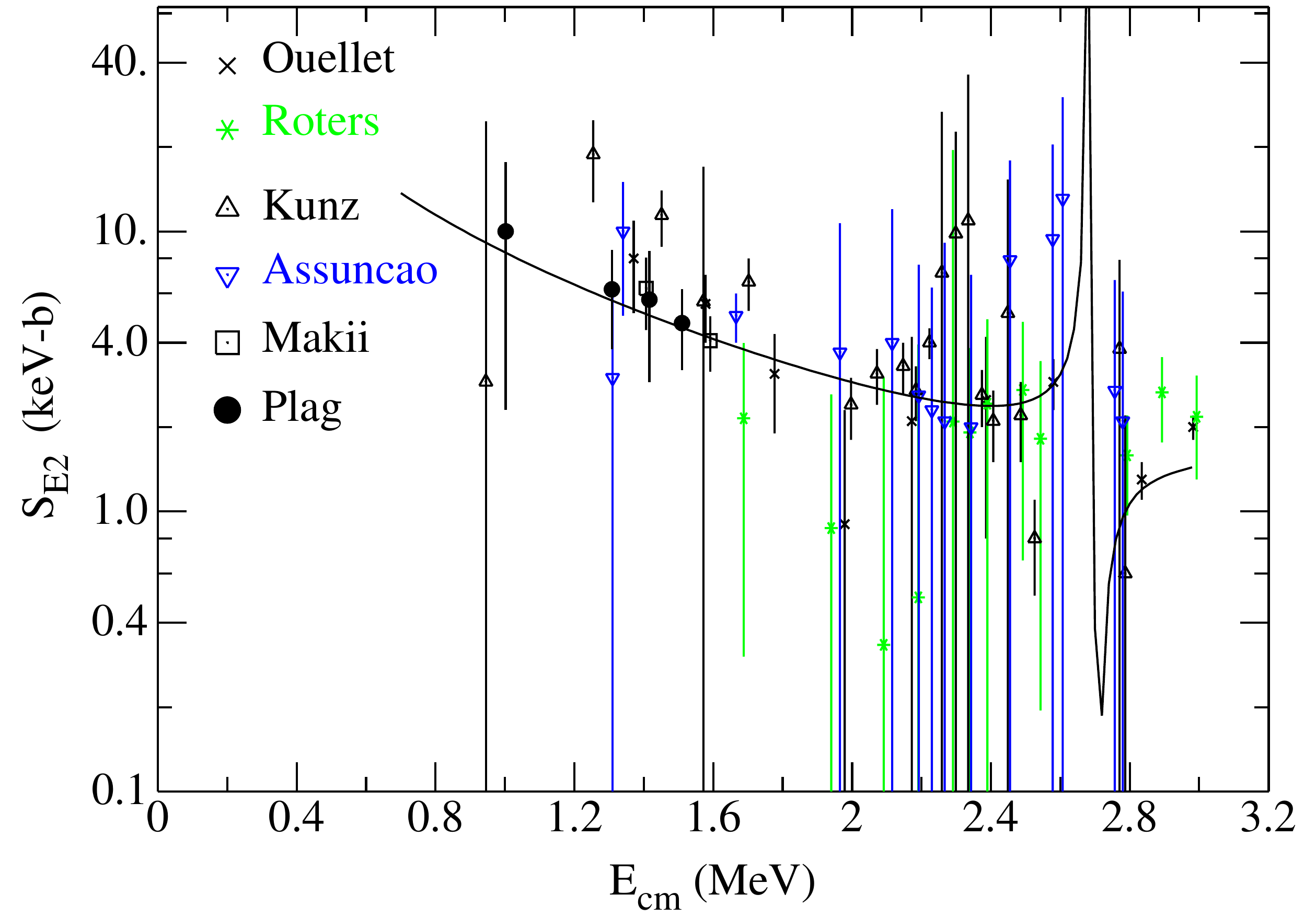}
\caption{The astrophysical S factor for the E1 (E2) cross section as a function of center of mass energy is shown in the left (right) panel.  The solid curves represent a typical fit from the parameters given in Table~\ref{two} and the data are taken from refs.~\cite{Dyer:1974pgc,Kremer:1988zz,Redder:1987xba,Ouellet:1992zz,Roters:1999zz,Gialanella:2001ayx,Kunz:2001zz,Assuncao:2006vy,Makii:2009zz,Plag:2012zz}.}
\label{fig1}
\end{figure}

\begin{table}[h] %add [H] placement to break table across pages
\caption{\label{two}Parameters used in the present simultaneous fits to original data for E1 and E2 for the curves shown in Fig~\ref{fig1}.  These parameters correspond to the ``all" fit in table~\ref {three}.  The widths for the bound states are reduced widths. The values marked with an asterisk were allowed to vary in the fit.  All other parameters were fixed.}
\begin{tabular*}{6.0in}{@{\extracolsep{\fill}}c c c c c c c}
\hline\hline
   &    &    E1   &    &         &   E2  & \\
$\lambda$ &  E$_\lambda$ (MeV) & $\Gamma_{\alpha_\circ}$ (keV) & $\Gamma_{\gamma_\circ}$ (eV) &E$_\lambda$ (MeV)&  $\Gamma_{\alpha_\circ}$ (keV) & $\Gamma_{\gamma_\circ}$ (eV) \\
\hline
1  &   -0.335 &  132.1$^*$            &  0.055          &   -0.448    &   89.9$^*$    &  0.097  \\
2  &    2.416  & 409.4$^*$            &  -0.0150$^*$    &    2.683    &   0.62            &   -0.0057\\
3  &    5.298  &  99.2                  &  5.6              &    4.407      &   83.0          &  -0.65  \\
4  &    5.835  &  -29.9                 &  42.0                &    6.092        &   -349          &  -1.02$^*$  \\
5  &   10.07     & 500                    &  0.604$^*$        &     -           &    -               &   -   \\
\hline\hline
\end{tabular*}
 \end{table}

The proposed JLab experiment\cite{suleiman:2014aa}} is expected to have several orders of magnitude improvement in luminosity over previous experiments and should provide data at the lowest practical values of energy.  The best R-matrix fit of the E1 and E2 data was taken as the most probable description of the projected JLab data. These JLab pseudo data were then randomly varied based on their projected uncertainties according to a Gaussian probability distribution about the best fit S-factor values. 
In order to study the impact of proposed JLab data and low energy data in general, three fits were performed:  a fit to existing E1 and E2 data (denoted by ``all" in table~\ref{three}), a fit to existing data and projected JLab data (denoted by ``all J'' in the table),  and a fit to all data in Fig.~\ref{fig1} above 1.6 MeV (denoted by ``E$>$1.6" in the table). Since the projected JLab data will be below 1.6~MeV, the cutoff of data below 1.6 MeV will give an indication of the importance of the lower energy data.  The S factors projected to 300 keV along with standard deviations, which represent the statistical fit uncertainty, are given in table~\ref{three} for the three cases.  The reduced $\chi^2$ for the fit to the original data is also shown.   As a test of the method, the error bars for the projected JLab data were arbitrarily reduced by an order of magnitude and the results are listed as ``all J/10" in the table.  Several observations can be made from the table.  The first is that the standard deviation for the total projected S-factor with proposed JLab data is smaller than that without JLab data.  Secondly, the total and E1 projections appear to be significantly larger for E$>$1.6 MeV data than the fits to ``all" data.   Thirdly, as expected the standard deviations for the ``all J/10" case are significantly smaller than that for the other cases.  Finally, the S-factor projections for E2 appear to be about a third of those for E1.

\begin{table}[h] %add [H] placement to break table across pages
\caption{\label{three}S-factor projections to 300 keV and standard deviations for total, E1 and E2.}
\begin{tabular*}{6.0in}{@{\extracolsep{\fill}} l c c c  c c c c}\hline\hline

data & orig $\chi^2$ & S  & $\sigma$  & S$_{E1}$  & $\sigma_{E1}$  &  S$_{E2}$  & $\sigma_{E2}$  \\ \hline

all            &    2.3      & 114.7           &  7.4        &   84.3         &     6.8        &    30.4     &   2.5  \\
all J          &    2.3       & 117.4            &  5.7        &   84.1       &    5.6       &     33.7    &    2.3  \\
all J/10    &     2.5      &  119.9   &        2.1            &  83.3        &    2.7      &      36.5    &   1.1   \\
E$>$1.6   &    2.6       &  132.8            &  4.2       &    100.6       &    3.6        &    32.2    &   2.5   \\
\hline\hline
\end{tabular*}
%\end{ruledtabular}
 \end{table}

\begin{figure}[t]
\begin{center}
\includegraphics[width=5.5in]{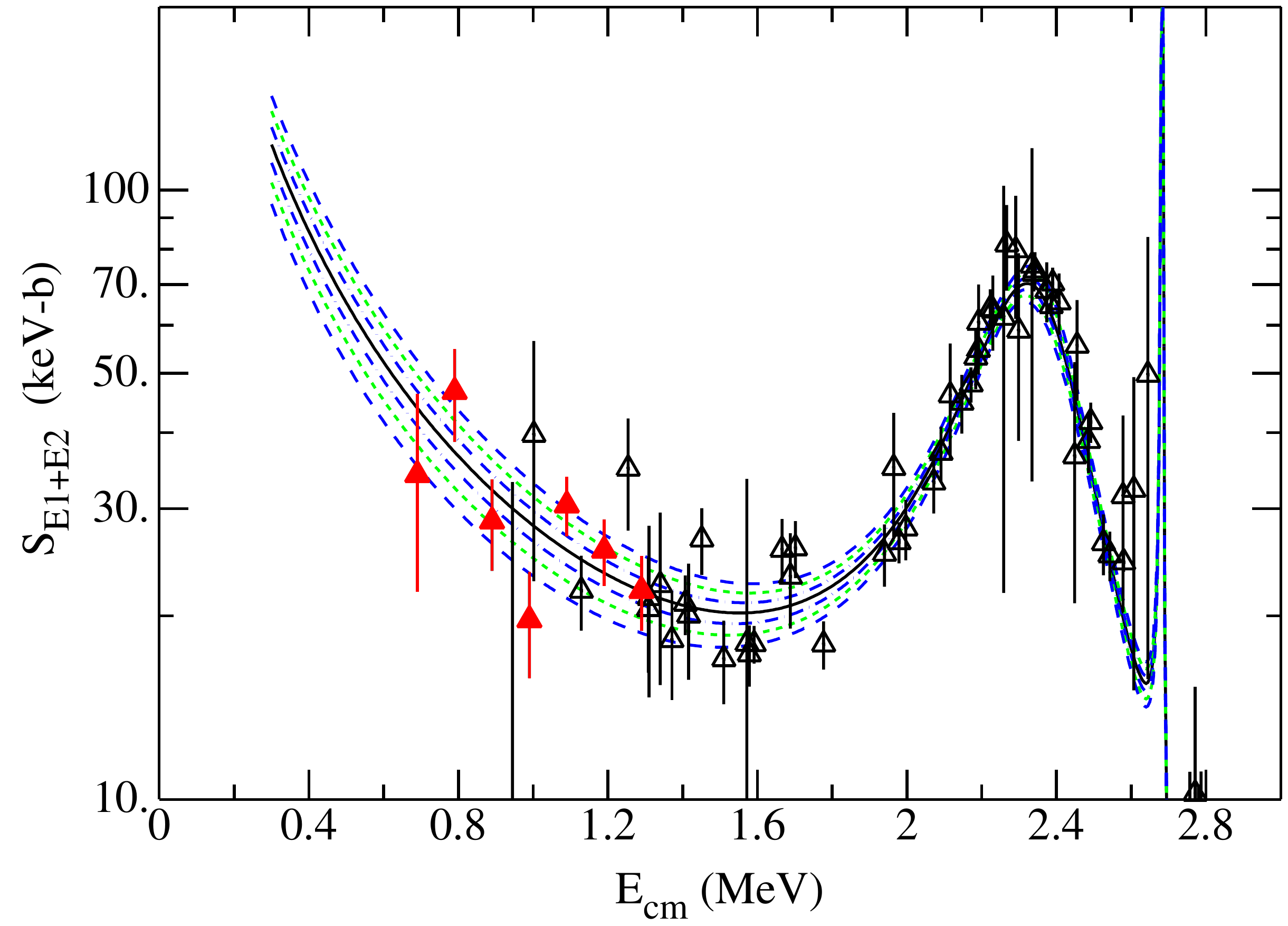}
\caption{Energy dependence of a fit to E1 and E2 data above 1.6~MeV indicating the $\pm$ 1, 2 and 3 standard-deviation bands shown as the dash-dot, short dash and long dash curves, respectively.  The open triangles represent a sum of E1 and E2 where both E1 and E2 data exist.  The standard deviation at 300~keV is given by the first line and fourth column of table~\ref{three}.  The projected JLab data are represented by the red triangles.}
\label{fig3}
\end{center}
\end{figure}

Fig.~\ref{fig3} shows curves that represent $\pm$ 1,2 and 3 standard deviation simultaneous fits to existing E1 and E2 data.  The curves are generated by performing 500 fits to the data, generating 500 sets of parameters similar to those in table~\ref{two}, and then using the parameter sets to determine the standard deviation at each value of energy.  The two and three standard deviation curves were estimated by simply multiplying the one standard values by factors of two and three, respectively.  The representative capture data, shown as open triangles, were taken as the sum of E1 and E2 results governed by where both E1 and E2 data exist.  The projected JLab data are represented by red triangles in the figure.  Given the statistical errors for the projected JLab data and the small number of values, one might not expect the projected JLab data to have a large impact on the statistical error.  This figure illustrates the importance of providing new data with significantly smaller errors at lower energy.

\section{Summary}

The projected JLab data were taken to represent E1 + E2 data since only total cross sections to the ground state will be measured.  The projected standard deviation for the 500 fits to the E1 and E2 data with the proposed JLab data is smaller than that without JLab data.   The JLab data constrain the total E1 + E2 cross section in the fit.  This leads to a smaller standard deviation than fitting E1 and E2 separately.  Fitting only data above 1.6~MeV leads to a significant shift upward in the projected S-factor at 300~keV.  This illustrates the importance of lower energy data in the extrapolation to 300~keV. Since the JLab experiment will provide new data below 1.6~MeV and even below existing data, we can infer that the proposed JLab data will have a significant impact on the value of the low energy extrapolation.  

%\newpage

\Acknowledgements

We thank O. Kirsebom for useful discussions.  This work is supported by the U.S. National Science Foundation under grant 1506459. This work is supported by the U.S. Department of Energy (DOE), Office of Science, Office of Nuclear Physics, under contract No. DE-AC02-06CH11357.

%\newpage
\bibliographystyle{unsrt}
\bibliography{CIPANP_Holt_final}

\end{document}